\documentclass[twocolumn,aps,prb]{revtex4}
\usepackage{amsmath,empheq}
\usepackage{amssymb}
\usepackage{mathrsfs}  
\usepackage{amsfonts}
\usepackage{booktabs}
\usepackage{physics}
\usepackage{graphicx} 
\usepackage{subfigure} 
\usepackage{epsfig}
\usepackage{color}
\usepackage{epstopdf}
 \usepackage{rotating}
 \usepackage{appendix}
 \usepackage[colorlinks=true,linkcolor=blue,citecolor=blue,filecolor=blue,urlcolor=blue]{hyperref}

\begin{document}
\bibliographystyle{prsty}

 \title{From quantum anomalous Hall phases to topological metals in interacting decorated honeycomb lattices}
 \author{Manuel Fern\'andez L\'opez and Jaime Merino}
\affiliation{Departamento de F\'isica Te\'orica de la Materia Condensada, Condensed Matter Physics Center (IFIMAC) and
Instituto Nicol\'as Cabrera, Universidad Aut\'onoma de Madrid, Madrid 28049, Spain}
\begin{abstract}
An analysis of the stability of topological states induced by Coulomb repulsion on decorated honeycomb lattices is presented. Based on a mean-field treatment of a spinless extended Hubbard model on the decorated honeycomb lattice we show how the quantum anomalous Hall (QAH) phase is a robust topological phase which emerges at various electron fillings and involves either quadratic band crossing points (QBCP) or Dirac points of the bands. The topological QAH phase is also found to be most stable against thermal fluctuations up to moderate temperatures when the Coulomb repulsion is maximally frustrated and at half-filling. We show how a topological metal can be induced from the QAH for certain electron doping ranges. Electrons on the Fermi surface of such metallic states are characterized by having non-zero Berry phases which can give rise to non-quantized intrinsic Hall conductivities.
%which could be measured experimentally. 
\end{abstract}
 \date{\today}

 \maketitle
 
\section{Introduction}
Topological insulators are being intensively studied since their discovery \cite{Discovered}. In spite of having a bulk gap, they are conducting in the surface with degenerate edge states crossing the Fermi energy which are protected by time-reversal symmetry (TRS). The associated topological bulk invariant, $Z_2$, characterizes the Quantum Spin Hall (QSH) phase predicted in graphene which, however, has too weak spin-orbit coupling (SOC).\cite{KaneMele, Z2} Stronger spin-orbit coupling (SOC) can be achieved in, for instance, 2D organic topological insulators (OTIs) based on honeycomb organometallic frameworks.\cite{Wang2013noint,khosla}  % Nevertheless, in recent works topological phases in systems with broken TRS have been discovered. 
In systems with broken TRS, the Quantum Anomalous Hall (QAH) phase which is characterized by the Chern number\cite{Fockphase}, can arise. This has been predicted, for instance, in OTIs in which TRS is broken by the strong magnetization field of the transition metal Mn ions.\cite{Wang2013int} Very recently, both kinds of topological phases have been theoretically predicted in the decorated honeycomb lattice (DHL). In a tight-binding model of the DHL in the presence of SOC, the QSH phase, characterized by a non-vanishing $Z_2$ invariant, emerges.\cite{nonint} When the Coulomb interaction is added to the tight-binding model, TRS is spontaneously broken giving way to a QAH phase characterized by a non-zero Chern number.\cite{int} This phase can be associated with the spontaneous generation of finite 'magnetic' fluxes piercing the elementary hexagonal placquettes of the lattice but with zero net flux through the unit cell. This phase is analogous to the quantum Hall phase without an applied magnetic field generated by adding complex next-nearest-neighbors (n.n.n) hopping amplitudes to the tight-binding model on the honeycomb lattice.\cite{haldane1988}

\begin{figure}[!t]
   \centering
\subfigure[]{\includegraphics[width=4.5cm]{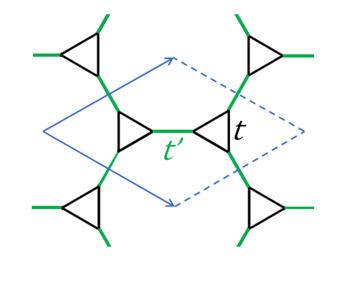}}
\subfigure[]{\includegraphics[width=3.2cm]{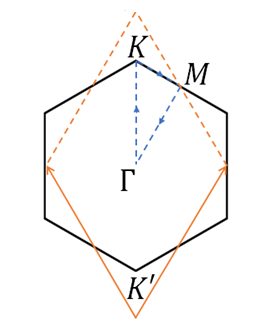}} 
\subfigure[]{\includegraphics[width=4.54cm]{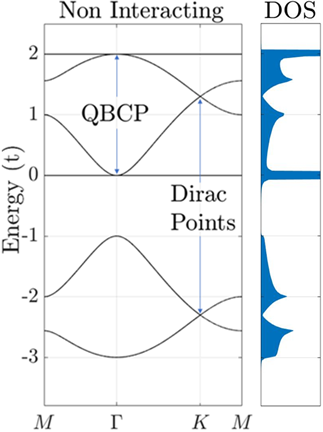}}
\subfigure[]{\includegraphics[width=3.82cm]{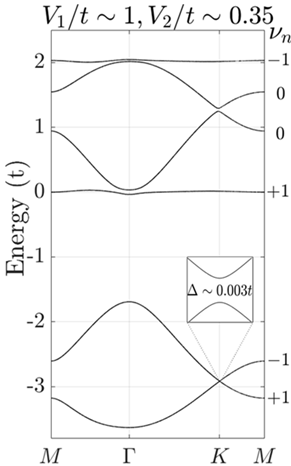}}   
       \caption{Band structure of the decorated honeycomb lattice. \small{(a) Unit cell and hopping parameters ($t$, $t'$) of the tight-binding model on a DHL,
       (b) the first Brillouin zone of the DHL, (c)
       the band structure of the tight-binding model on a decorated honeycomb lattice for $t=t'$ is shown along the $k$-path $M\rightarrow \Gamma \rightarrow K \rightarrow M$. The corresponding density of states is plotted at the side. The band structure displays Dirac points at $K$ (and $K'$) 
       %at the Fermi energy for filling fractions $f=1/6$ and $f=2/3$, 
       as well as quadratic band crossing points (QBCP) at the $\Gamma$-point 
       %at $f=1/2$ and $f=5/6$.
       (d) The Hartree-Fock energy bands of the spinless model (\ref{Hubbard}) in the presence of the Coulomb interaction ($V_1\sim 1, V_2/t \sim 0.35$ at $f=1/2$). Topologically non-trivial gaps open up at the $\Gamma$ point and $K$ ($K'$) points 
       making the system insulating at integer fillings. The Chern number of each band is shown on the side of the plot. When 
       doping with electrons off half-filling between $f=1/2$ and $f=2/3$ topological metallic states arise. Inset of Fig. \ref{fig:fig0}\textcolor{blue}{(d)} shows the magnitude of the gap at the $K$ point.
       %It is possible to calculate the Chern number ($\nu_n$) associated with each band $n$. 
       %The two lowest energy bands have similar topological properties to the graphene bands with ($\nu_{1,2}=\pm 1$). However, the $f=2/3$ and $f=5/6$ bands have $\nu_{4,5}=0$ due to Berry phase cancellation between $\Gamma$ (QBCP) and $K$ ($K'$) (Dirac points).}}
      %{\color{red} Red and green lines correspond to hoppings with phases 0 or $\pi$, respectively.
      }}
 \label{fig:fig0}
 \end{figure}

The DHL is interesting not only from the theoretical point of view but also because it is realized in actual materials such as the trinuclear 
organometallic compounds\cite{khosla,jacko,merino,powell} {\it e. g.}  Mo$_3$S$_7$(dmit)$_3$, in Iron (III) acetates\cite{iron3} or in cold fermionic atoms loaded in a decorated honeycomb optical lattice\cite{cdmft}. The hopping parameters entering the tight-binding hamiltonian (\ref{Hubbard}), $\mathcal{H}_{tb}$, are shown in Fig. \ref{fig:fig0}\color{blue}(a)\color{black}. Hence, the DHL can be seen as interpolating between the honeycomb and the Kagom\'e lattice.\cite{jacko1} The band structure is richer than on the honeycomb lattice potentially leading to novel topological states of matter. Apart from the Dirac points protected by TRS and inversion symmetry (IS), the band structure also displays quadratic band crossing points (QBCP) which are topologically protected by time-reversal invariance and $C_4$ or $C_6$ symmetries.\cite{Sun} Unlike Dirac points carrying Berry fluxes of $\pm \pi$, QBCPs can carry Berry phases of $\pm 2\pi$ leading to non-trivial topological phases. For instance, a robust QAH phase is induced by Coulomb repulsion in the checkerboard lattice containing a QBCP protected by $C_4$ symmetry.\cite{Zeng2018} It is then 
interesting to search for topological phases associated with the QBCP's on the DHL 
shown in Fig. \ref{fig:fig0}{\color{blue}(c)} which respect the C$_6$ symmetry of the
lattice. These QBCP leads to divergences in the density of states (DOS) which are relevant for inducing instabilities, particularly around $f=1/2$ and $f=5/6$. 

Non-trivial topology arises when Coulomb repulsion acts between electrons
on the DHL. The off-site Coulomb repulsion in the spinless extended Hubbard  
model (\ref{Hubbard}) can lead to gaps at $\Gamma$ and $K$ ($K'$) so that the
system becomes an insulator. Such insulator is topologically non-trivial
consisting on some bands with non-zero Chern numbers ($\nu_n$) as displayed in Fig. \ref{fig:fig0}\color{blue}(d)\color{black}, implying topologically non-trivial states arising at certain electron fillings. These states are due to the spontaneous formation of QAH phases induced by the Coulomb repulsion.\cite{int} The two lowest energy bands with Chern numbers: $\nu_{1,2}=\pm 1$ closely resemble those of the QAH found in a honeycomb lattice when sufficiently strong off-site Coulomb repulsion is considered.\cite{Fockphase} In contrast, the bands $n=4,5$ are effectively topologically trivial, $\nu_{4,5}=0$, due to the cancellation of the Berry phase (of $\pm 2 \pi$) involving the QBCP around the $\Gamma$-point and the Berry phases of $\pm \pi$ associated with the Dirac cones at $K$ and $K'$.
 
In the present work, we study the doping and temperature effects on the QAH phase generated by Coulomb repulsion on a DHL. Previous work on the DHL has concentrated on the non-interacting QSH phase induced by SOC and the QAH phase induced by Coulomb repulsion but only at half-filling, $f=1/2$. However, little is known about the existence of the QAH at other fillings and/or as temperature is raised. We cover this gap by obtaining the full phase diagram of an spinless extended Hubbard model with nearest-neighbors (n.n), $V_1$, and n.n.n, $V_2$ Coulomb repulsion on the DHL at all relevant integer fillings and at certain non-integer fillings. We find topological metallic phases arising at filling fractions varying, for instance, between $f=1/2$ and $f=2/3$ by 
fixing the ($V_1,V_2$) parameters for which the QAH is the ground state. Non-zero intrinsic Hall conductivities as well as magnetoresistance oscillation experiments associated with the Berry phases on the Fermi surface can provide experimental evidence for the existence of such topological metals. 
%The Fermi level is displaced in such a way that it can lie either on a gap zone, finding insulating phases, or on a gapless one, finding metallic phases. We consider an spinless extended Hubbard model with first ($V_1$) and second ($V_2$) neighbour potential parameters.

The paper is organized as follows. In Sec. \ref{sec:methods} we describe the model and the Hartree-Fock method used to analyze the model. In Sec. \ref{sec:results} we obtain the ground state phase diagram of the model at the filling fractions, $f$, at which either QBCP or Dirac points are relevant {\it i. e.} for $f=1/6$, $f=1/2$, $f=2/3$ and $f=5/6$. We also explore in Sec. \ref{sec:results} the thermal stability of the QAH phase found at $T=0$ providing a $T-V$ phase diagram for specific Coulomb parameters for which the QAH is the ground state.
By electron doping the QAH we show how topological metallic phases 
arise. Finally, in Sec. \ref{sec:discussion} we discuss the implications of our findings on experimental observations of the Hall conductivity and magnetoresistance oscillations. We close up the paper with the conclusions in Sec. \ref{sec:concl}.

\section{Methods}\label{sec:methods}
In order to analyze possible topological states emerging from the Coulomb repulsion, we consider a spinless extended Hubbard model on a DHL:
\begin{align}
\left.\begin{aligned}
\mathcal{H} &= \mathcal{H}_{tb}\qquad+\qquad\mathcal{H}_{Coul} \\
\mathcal{H}_{tb} &=-t \sum_{\langle ij\rangle\bigtriangleup }c_{i}^{\dagger}c_{j}-t' \sum_{\langle ij\rangle\bigtriangleup\rightarrow\bigtriangleup}c_{i}^{\dagger}c_{j} \\
\mathcal{H}_{Coul} &= V_1\sum_{\langle ij\rangle}n_{i}n_j+V_2\sum_{\langle\langle ij\rangle\rangle}n_in_j,
\end{aligned}\quad\right.
\label{Hubbard}
\end{align}
where the fermion occupation operator is defined as $n_i\equiv c^\dagger_ic_i$. We consider the Coulomb repulsion between electrons in nearest and next-nearest 
neighbor sites parameterized by $V_1$ and $V_2$, respectively. 
We have neglected the n.n.n hopping amplitudes as suggested by {\it ab initio} DFT calculations on \cite{jacko1} Mo$_3$S$_7$(dmit)$_3$ crystals.
Since this hamiltonian is cuartic in the fermion operators, it cannot be solved exactly so we apply a Hartree-Fock mean-field decoupling of these terms:
\begin{equation}
n_in_j\sim (n_in_j)_{\text{Hartree}} - (n_in_j)_{\text{Fock}} 
\end{equation}  
where $(n_in_j)_{Hartree}=n_i \langle n_j\rangle +  \langle n_i\rangle n_j-\langle n_i \rangle\langle n_j\rangle $ and $(n_in_j)_{Fock}=c_i^\dagger c_j\langle c_j^\dagger c_i\rangle+\langle c_i^\dagger c_j\rangle c_j^\dagger c_i -\langle c_i^\dagger c_j\rangle\langle c_j^\dagger c_i\rangle$. We work in the canonical ensemble with a fixed number of electrons $N_e$. At a given temperature $\frac{1}{\beta}=k_B T$, the free energy $\mathcal{F}$ is given by $\mathcal{F}=\mathcal{F}_T+\mathcal{F}_H+\mathcal{F}_F$, where:
\begin{align}
\label{FreeT}
&\mathcal{F}_T=-k_BT\sum_{\textbf{k},n}log[1+e^{-\beta(E_{\textbf{k},n}-\mu)}]+\mu N_e&  \\
\label{FreeH}
&\mathcal{F}_H=-{V_1}\sum_{\langle ij\rangle}\langle n_i \rangle\langle n_j\rangle-V_2\sum_{\langle\langle ij\rangle\rangle}\langle n_i \rangle\langle n_j\rangle& \\
\label{FreeF}
&\mathcal{F}_F=V_1\sum_{\langle ij\rangle}\langle c_i^\dagger c_j\rangle\langle c_j^\dagger c_i\rangle + {V_2}\sum_{\langle\langle ij\rangle\rangle}\langle c_i^\dagger c_j\rangle\langle c_j^\dagger c_i\rangle&
\end{align}
with $\mu$ the chemical potential and $E_{\textbf{k},n}$ the Hartree-Fock band
dispersions. In order to find the mean field amplitudes that minimize the free energy we solve the following system of coupled equations:
\begin{align}
\label{HarS}
\frac{\partial \mathcal{F}}{\partial \langle n_i\rangle}&=0\\
\label{Fock1}
\frac{\partial \mathcal{F}}{\partial \langle c_i^\dagger c_j\rangle}&=0\\
\label{Fock2}
\frac{\partial \mathcal{F}}{\partial \langle c_j^\dagger c_i\rangle}&=0
\end{align}
The number of variables of the whole system is 27 (6 $\langle n_i\rangle$ and 21 $\langle  c_i^\dagger c_j\rangle$). Since finding the global minimum of $\mathcal{F}$ is not a straightforward task, we first fix $V_1\sim V_2\ll t$ and solve the equations at the filling fraction, $f=1/2$. We can see that the converged solutions  display the following pattern in the Fock terms:  
\begin{align}
\left.\begin{aligned}
\langle  c_i^\dagger c_j\rangle&\equiv \chi_1=\xi_1+i\eta_1 \qquad \bigtriangleup  &1^{st} \\
\langle  c_i^\dagger c_j\rangle&\equiv \chi_2=\xi_2 +i\eta_2 \quad \bigtriangleup\rightarrow\bigtriangleup  &1^{st} \\
\langle  c_i^\dagger c_j\rangle&\equiv \chi_3=\xi_3 +i\eta_3 &2^{nd}
\end{aligned}\quad\right.
\label{ansatz}
\end{align}
where $\bigtriangleup$ and $\bigtriangleup\rightarrow\bigtriangleup$ mean the intratriangle and intertriangle neighbours, respectively. In actual calculations, we reduce our system of equations by fixing this $ansatz$. We observe that the Fock amplitudes that minimize the energy are complex $\chi_m\in \mathbb{C}$ and the mean densities at each site $\langle n_i\rangle$ are all the same at this filling. This means that time-reversal symmetry is spontaneously broken while the rotational $C_6$ symmetry is preserved. This phase is a quantum anomalous Hall (QAH) phase and it is characterized by the presence of finite fluxes ($\eta_m$) through elementary placquettes of the lattice (see Fig. \ref{scheme}\color{blue}(b)\color{black}). However, the total flux through the unit cell is zero due to the periodic boundary conditions. The Fock contribution to the total Hartree-Fock free energy $\eqref{FreeF}$ assuming this $ansatz$ takes the form:
\begin{align}
\mathcal{F}_{F}=6V_1(\xi_1^2+\eta_1^2)+3V_1(\xi_2^2+\eta_2^2)+12V_2(\xi_3^2+\eta_3^2)
\label{freeFA}
\end{align}
Looking at the free energy expressions $\eqref{FreeH}$ and $\eqref{freeFA}$, the equations for the densities $\eqref{HarS}$ and for the Fock amplitudes $\eqref{Fock1}$ seem to be decoupled. However, due to the dependence of the band dispersions $E_{\textbf{k},n}(\langle n_i \rangle, \xi_m,\eta_m)$ on both the local densities and the Fock amplitudes entering the thermal part $\eqref{FreeT}$, they form a set of coupled self-consistent equations:
\begin{align}
\left.\begin{aligned}
V_1\sum_{j \in \langle ij\rangle} n_j+V_2\sum_{j \in\langle\langle ij\rangle\rangle} n_j=\sum_{\textbf{k},n}\frac{\partial E_{\textbf{k},n}/\partial  n_i}{1+e^{\beta(E_{\textbf{k},n}-\mu)}}\\
\xi_m=-\frac{1}{a_mV_{1,2}}\sum_{\textbf{k},n}\frac{\partial E_{\bf{k},n}/\partial \xi_m}{1+e^{\beta(E_{\textbf{k},n}-\mu)}}\\
\eta_m=-\frac{1}{a_mV_{1,2}}\sum_{\textbf{k},n}\frac{\partial E_{\bf{k},n}/\partial \eta_m}{1+e^{\beta(E_{\textbf{k},n}-\mu)}}
\end{aligned}
\quad\right.
\label{HFEqs}
\end{align}
where $i$ is the sites index ($1\leq i\leq 6$) and $m$ is the Fock $ansatz$ index  ($1\leq m\leq 3$). The sums over $j\in\langle ij\rangle$ and $j\in\langle\langle ij\rangle\rangle$ refer to the n.n and n.n.n of each site $i$, respectively. $a_m$ are coefficients which come from the derivation of $\eqref{freeFA}$, and $V_{1,2}$ is either $V_1$ or $V_2$ for $m=1,2$ and $m=3$ respectively $\eqref{ansatz}$. The equations are written in a compact way: six equations in the first line corresponding to each site $i$ while in each resting ones (second and third line), three equations are shown coming from each $ansatz$ index, $m$. This system of 12 equations can be solved iteratively. 
%by plugging in the right side of the equations the solutions of the previous iteration.
 
Although the system has been simplified through the {\it ansatz} assumed, in some situations the convergence to one or other minima %is difficult and 
may depend on the initial guess seeded into the equations. Hence, we proceed as follows. For a given set of $(V_1,V_2)$, we search for the global minimum by comparing the free energy of the complete set of equations $\eqref{HFEqs}$ with the free energies obtained with the same set of equations satisfying additional constraints. In this way we can reduce the unwanted dependence of the solution on the initial guess plugged into the system of equations. The first constraint imposed on the equations consists on assuming a pure Hartree decoupling which fixes $\chi_m=0$, so that the system of equations is restricted to the $\langle n_i \rangle$ only giving the different possible charge ordering patterns of the model. The second constraint consists on imposing Fock amplitudes which are purely imaginary, $\xi_m=0$, 
giving the contribution to the chiral currents. The third and final constraint restricts the system to solutions of the type: $\eta_m=0$, so that the $\chi_m$, being real, describe %effective
shifts of the hopping parameters $\delta t_m$: $t(\bigtriangleup)\rightarrow -t+V_1\delta t_1$, $t(\bigtriangleup\rightarrow\bigtriangleup)\rightarrow -t'+V_1\delta t_2$, $t(2^{nd})\rightarrow 0+V_2\delta t_3$. As stated above, the free energies $\mathcal{F}$ are evaluated separately assuming each of the three constraints described above and compared to full set of equations without any constraint imposed. The solution with the lowest free energy provides the ground state for a fixed set of $(V_1,V_2)$. This allows to construct the phase diagrams of 
Fig. \ref{fig:fig1}.

The topological properties of the insulating solutions are characterized by calculating the total Chern number ($\nu$) of the system. The QAH phase is characterized by having a non-zero Chern number ($\nu\neq 0$) in analogy with the standard Quantum Hall effect as shown in Fig. \ref{scheme}\color{blue}(b)\color{black}. The Chern numbers are evaluated numerically by evaluating the Berry flux through the elementary placquettes in which the first Brillouin zone is discretized which neutralizes the arbitrary phase coming from the gauge invariance. The topological properties of metallic states are analyzed in a similar way by computing the Berry flux through the $FS$ defined by the partially filled bands. This gives the non-quantized contribution to the intrinsic Hall conductivity $\sigma_{xy}$.\cite{Haldane} The details regarding these numerical procedures are given in Appendix A.

%From the computation of the free energies and the Chern number at a given $V_1$, $V_2$ we have obtained the ground state, being able to construct a phase diagram. We get one for each filling fraction $f$ involving at each case different kinds of touching points: either QBCP or Dirac points.

\section{Results}
\label{sec:results}

 In this section, we first explore the ground state of the model for different $(V_1,V_2)$ following the procedure described above at each filling fraction, $f$. Then we check the stability of the QAH phase against thermal fluctuations. This is done for the fixed $V_2/V_1$ ratio at which the QAH is the ground state of the model. We solve the equations at finite temperature, obtaining a $T-V$ phase diagram. This allows to extract an estimate of the temperature below which the QAH phase is stable. Finally, 
 we electron dope the system varying the filling between $f=1/2$ and $f=2/3$ within the $V_1,V_2$, at which the QAH state is the ground state of the system. In this way we study the evolution of the Fermi surface and the possible topological metallic states emerging in the model.
 
\subsection{Ground state phase diagram} 

\begin{figure}[!t]
   \centering
\includegraphics[width=4.25cm]{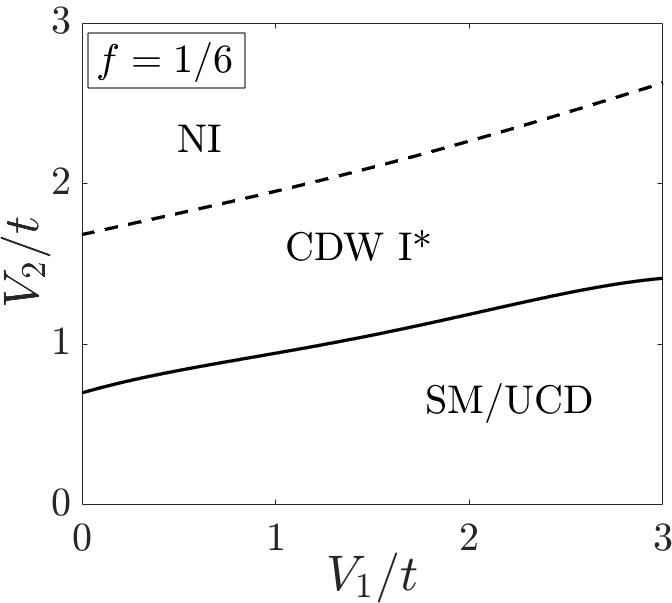}
\includegraphics[width=4.25cm]{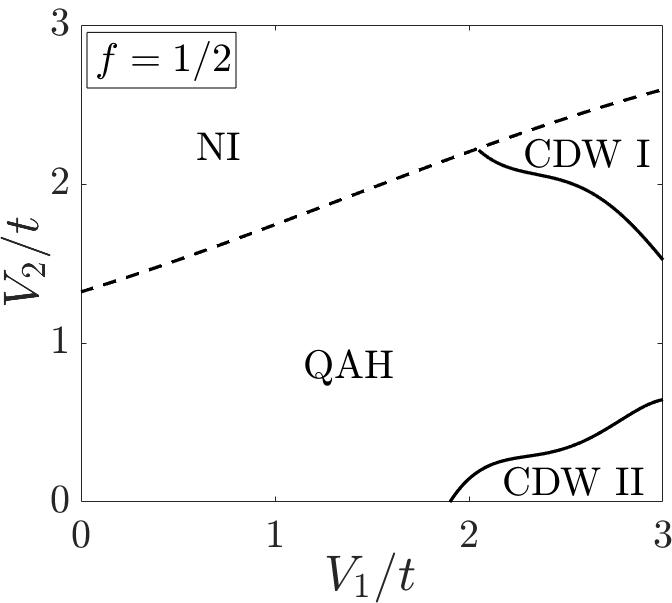}
\includegraphics[width=4.25cm]{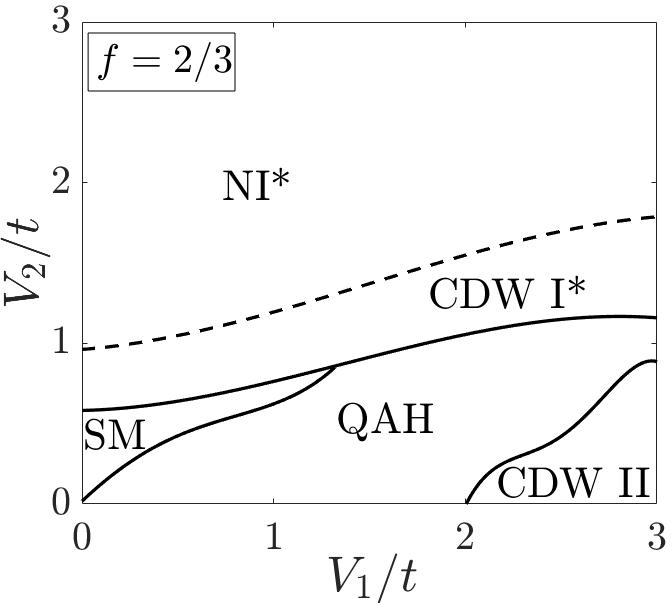}
\includegraphics[width=4.25cm]{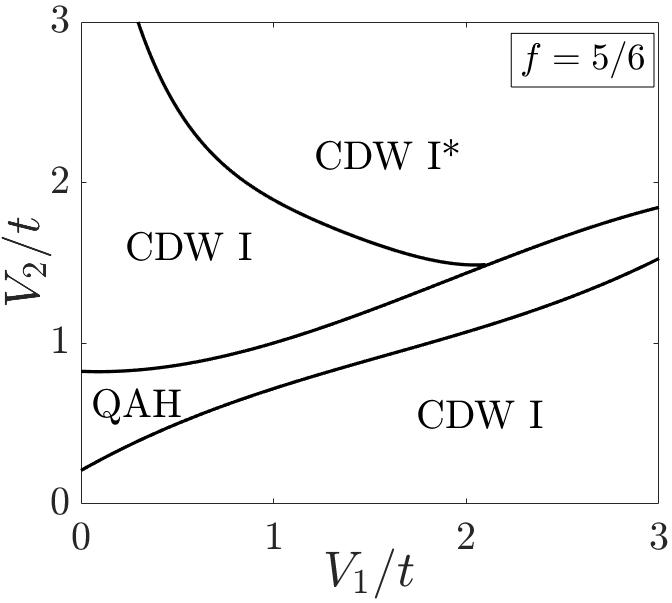}
       \caption{\small{The $V_1-V_2$ phase diagrams at various filling fractions, $f$. (a) At $f=1/6$ involving Dirac points: semimetalic (SM) and nematic insulating (NI) phases arise. (b) At $f=1/2$ with a QBCP: a quantum anomalous Hall (QAH) phase, two types of charge density wave phases, CDW I and CDW II as well as a NI phase emerge. (c) $f=2/3$ with Dirac points: a SM phase which becomes unstable to a QAH phase as well as
       %Coulomb interaction is more relevant than at $f=1/6$. 
       charge ordered phases such as CDWI*, CDW II and NI* appear. (d) $f=5/6$ with a QBCP: a QAH phase
       %also emerges but in an smaller region due to the the weaker effect of the Coulomb interaction since there is only one hole per unit cell. 
       and charge ordered phases CDW I and CDW I* are found. Dashed lines denote second order transitions while full lines first order transitions.}}
 \label{fig:fig1}
 \end{figure}

We first discuss the ground state phase diagram of model (\ref{Hubbard}). We obtain the phase diagrams at four filling fractions, $f$, at which the conduction and valence bands have  either Dirac or quadratic band touching points. We search for
the ground state of the model for given $V_1$, $V_2$ following the method explained in the previous section (see Fig. \ref{fig:fig1}).
 
\begin{figure}[!t]
   \centering
\subfigure[]{\includegraphics[width=2.8cm]{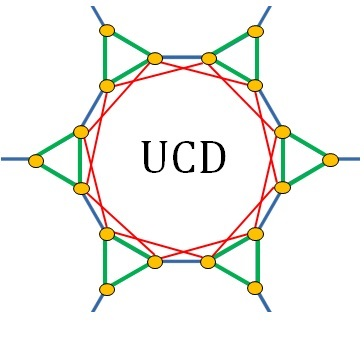}}
\subfigure[]{\includegraphics[width=2.8cm]{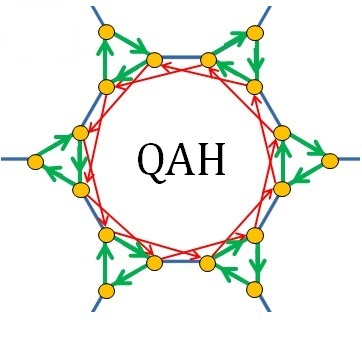}}
\subfigure[]{\includegraphics[width=2.8cm]{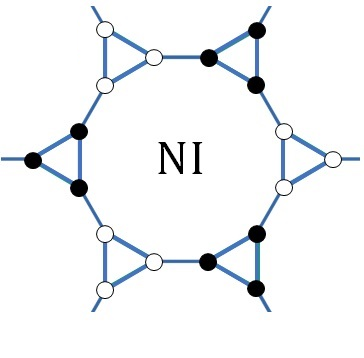}}
\subfigure[]{\includegraphics[width=2.8cm]{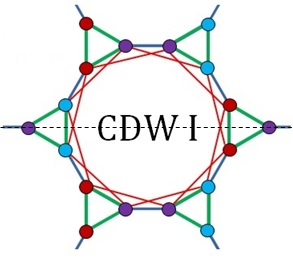}}
\subfigure[]{\includegraphics[width=2.8cm]{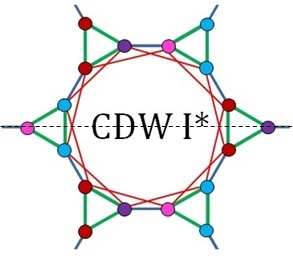}}
\subfigure[]{\includegraphics[width=2.8cm]{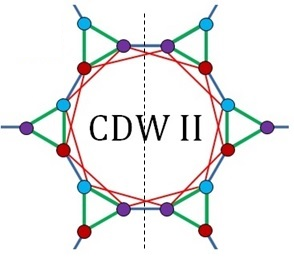}}
       \caption{\small{The different ground states shown in the $V_1-V_2$ phase diagrams of Fig. \ref{fig:fig1}{\color{blue}(a)}. Uniform charge density (UCD). The charge density is uniformly distributed in the unit cell and the bonds $\xi_m\neq 0$ follow the $ansatz$ \eqref{ansatz} preserving $C_6$ symmetry. (b) Quantum anomalous Hall (QAH) phase. This phase preserves $C_6$ symmetry while TRS is broken by spontaneous chiral currents $\eta_m\neq 0$ being two-fold degenerate depending on the direction of the current (the represented corresponds to a Chern number, $\nu= +1$). (c) Nematic insulator (NI). In this state the charge inside the unit cell is located in one of the triangles of the unit cell. The rotational symmetry is reduced from $C_6$ to $C_3$ leading to a two-fold degenerate ground state. (d) Charge density wave I (CDW I). The charge is distributed following the colour patterns displayed. Inside each unit cell nearest-neigbor sites are paired up with the same charge. (e) CDW I*. In this phase the inter-triangle nearest-neighbours sites have different densities. (f) Charge density wave II (CDW II). The densities are associated in pairs but not between nearest-neighbors (except for the n. n. intertriangle sites). In the CDW phases the hamiltonian is invariant under reflection transformations: with respect to the $x$-axis for CDW I and CDW I*, and with respect to the $y$-axis for CDW II. All these states are three-fold degenerate.}}
 \label{scheme}
 \end{figure}

The resulting phase diagrams are shown in Fig. \ref{fig:fig1}. We find uniform charge density (UCD), Charge Density Wave (CDW) and Nematic Insulator (NI) phases whose spatial patterns are displayed in Fig. \ref{scheme}. 
While the uniform charge density (UCD) preserves the symmetries of the lattice, the CDW phases break 
the $C_6$ rotational invariance reducing it to a mirror symmetry only. The reflection plane goes through either intra-triangle (in CDW I and CDW  I* phases) or inter-triangle 
(in CDW II) bonds.Since our phase diagram is obtained on the six site unit cell of the lattice only
$q=0$ CDW states are obtained. Increasing the size to 24 sites can lead to ${\bf q} \neq 0$ CDW phases around
the QAH phase. However,  
we find that the parameter region for which the QAH phase is stable is robust against increasing the lattice size independently of the 
CDW patterns stabilized.
The CDW states found are three-fold degenerate due to the invariance of the total energy against 60$^0$ rotations of the lattice.  

The NI phase consists of an
almost empty triangle and the rest of the charge uniformly distributed in the other triangle of the unit cell. This state is two-fold degenerate due to the reduction of rotational symmetry from $C_6$ down to $C_3$. The transition line separating this phase from the CDW or QAH phases is second order as shown in Fig. \ref{fig:fig1}. When $\sim 90\%$ of the total charge in a unit cell is localized in one of the triangles we assume that the transition has occurred. With four electrons per unit cell, $f=2/3$, three electrons are located in one triangle while the electron left
distributes uniformly among the sites of the other discharged triangle. We denote this phase by NI*.
Comparing the phase diagrams at different fillings, we conclude that the QAH phase is robust occurring at all filling fractions except for $f=1/6$. We can rationalize this from the 
fact that, at this filling, there is only one electron per unit cell so that the effective Coulomb repulsion is not strong enough to destabilize the semimetalic (SM) phase and turn it into the QAH phase. This has also been found in the Kagom\'e lattice, in which the emergence of the QAH phase requires a third 
neighbour interaction.\cite{int} However, at $f=5/6$ where we only have one hole per unit cell we do find a stable QAH region. This seems to be counter-intuitive if we replace in the above argument particles by holes. However, the different phase diagrams and the larger stability of the QAH at $f=5/6$ compared to $f=1/6$ can be attributed to the different density of states (DOS). In particular, the DOS at fillings involving QBCP's displays a divergence due to the flat bands as shown in Fig. \ref{fig:fig0}\color{blue}(c)\color{black}. Both uniform charge density (UCD) and QAH phases preserve $C_6$ rotational symmetry but the QAH breaks TRS $\eta_m\neq 0$, as shown in Fig. \ref{scheme}. At filling fractions involving QBCP's {\it i. e.} at $f=1/2$ and $5/6$ there are no regions UCD phases. Due to the divergence of the DOS, a way to destabilize the QAH phase is through breaking the spatial symmetry caused by charge ordering phenomena. 
We finally note that at $f=1/2$, along the line $V_1/t\sim V_2/t$, the QAH phase is the most stable. This means that the energy difference with the competing UCD state reaches its maximum at this filling (see Fig. \ref{comparison}{\color{blue}(a)} in Appendix B). Hence, we choose this range of parameters with $f=1/2$ in order to explore the stability of the QAH against thermal fluctuations. 

\subsection{Finite temperatures}

\begin{figure}[!b]
   \centering
    \includegraphics[width=7.5cm]{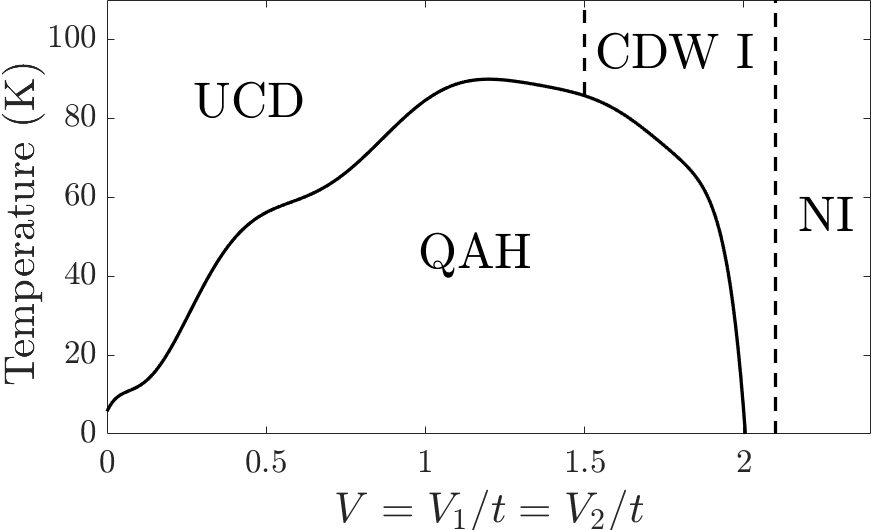}
       \caption{ $T-V$ phase diagram at filling fraction $f=1/2$. At zero temperature the QAH phase emerges along the line $V=V_1/t\sim V_2/t$ until $V\sim 2$ where charge order sets in destroying the topological phase. Temperature destroys the QAH phase without loosing $C_6$ symmetry except for $1.5< V< 2$ in which the transition to a CDW I takes place. We find a robust QAH phase up to temperatures of $T\sim 84$ K. The hopping of $t=0.05$ eV is taken for obtaining temperatures in Kelvin. }
      %{\color{red} Red and green lines correspond to hoppings with phases 0 or $\pi$, respectively.}
 \label{VT}
 \end{figure}

As stated above we analyze the effect of temperature on the QAH phase
at $f=1/2$ for $V\equiv V_1/t\sim V_2/t$. Our mean-field $T-V$ phase diagram is shown in Fig. \ref{VT}. In the phase diagram temperatures are given in Kelvin using
the hopping parameter $t=0.05$ eV corresponding to Mo$_3$S$_7$(dmit)$_3$ crystals.\cite{model} In the limit $T\rightarrow 0$ we do recover the results shown in Fig. \ref{fig:fig1}, as expected. The QAH state is the most stable phase when $V \lesssim 2$, above this value the transition to the CDW I occurs. Subsequently around $V\sim 2.1$, a second order phase transition to the NI phase occurs. Observe in the phase diagram how thermal fluctuations induce a transition from a QAH phase to a UCD for $0<V<1.5$ and 
to a CDW I for $1.5<V<2$. In the former case the $C_6$ rotational invariance of the lattice is preserved across the transition whereas the latter transition involves a spontaneous breaking of the rotational symmetry of the lattice. We do not find any further charge ordering transitions beyond the temperature range shown in the figure implying the robustness of the NI and CDW I against thermal fluctuations. 
From our phase diagram we conclude that the maximum temperature at which we find an stable QAH phase is: $T\sim 84$ K for $V_1\sim V_2\sim 1.2 t$. In Fig. \ref{comparison}{\color{blue}(b)} of appendix B we compare the free energies of the UCD and QAH phases. Based on our mean-field theory analysis we conclude that the QAH phase may be most likely found in half-filled isolated layers of Mo$_3$S$_7$(dmit)$_3$ in a broad range of temperatures
if the $V_1$ and $V_2$ parameters are tuned through the optimal $V_1\sim V_2\sim 1.2 t$ values. However, it remains to be seen whether other phases different to the QAH phase \cite{cdmft} become the ground state.
We note that the temperature scale of 84 K found here is readily consistent with the large excitation gap of about $0.18 t$ corresponding to 100 K 
found for $V_1\sim V_2\sim 1.2 t$. The large temperature scale found for the stability of the QAH phase suggests that decorated honeycomb lattice materials are 
good candidates for hosting QAH phases.

\subsection{Topological metals}

We now explore the possibility of stabilizing a topological metal by doping the QAH state. By raising the Fermi level above zero we partially fill the fourth band in Fig. \ref{fig:fig0}{\color{blue}(b)} without closing the band gaps. This leads to a topological metallic state with broken TRS since the chiral currents giving rise to the QAH at $f=1/2$ are found to persist even at non-integer fillings between $f=1/2$ and $f=2/3$ and between $f=2/3$ and $f=5/6$. 

\begin{figure}[!b]
   \centering
\includegraphics[width=2.7cm]{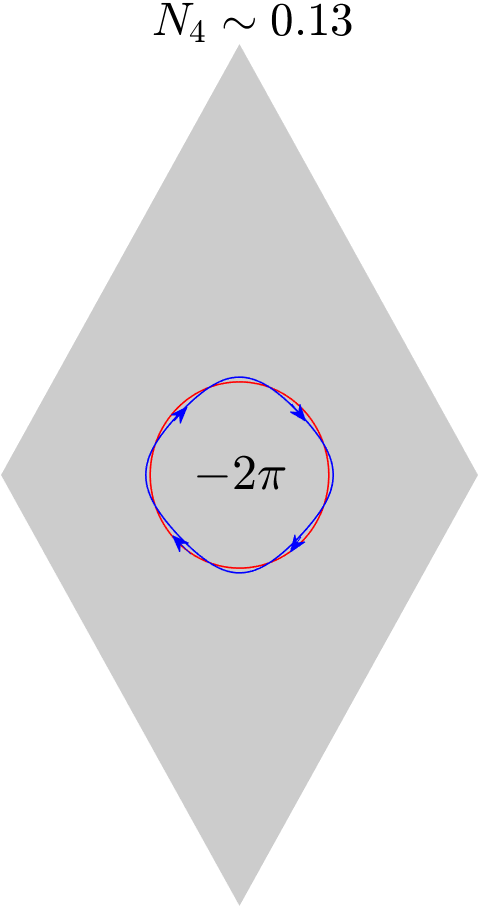}
\includegraphics[width=2.7cm]{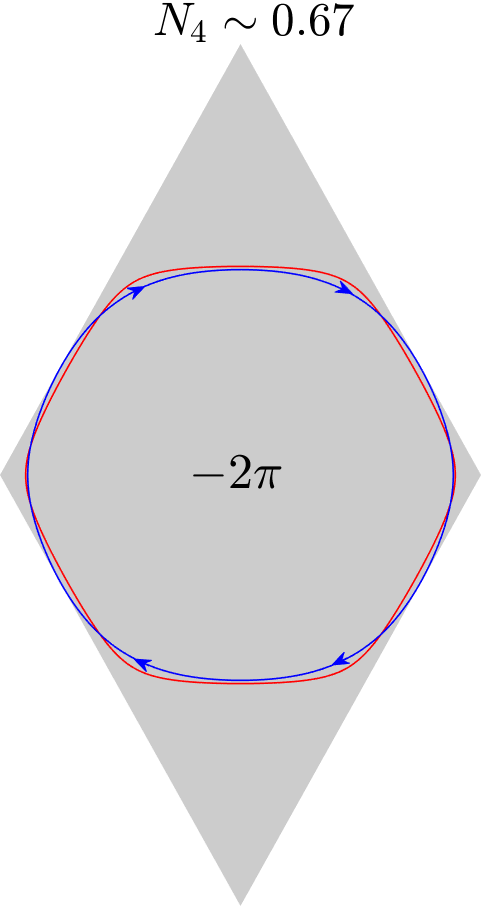}
\includegraphics[width=3.cm]{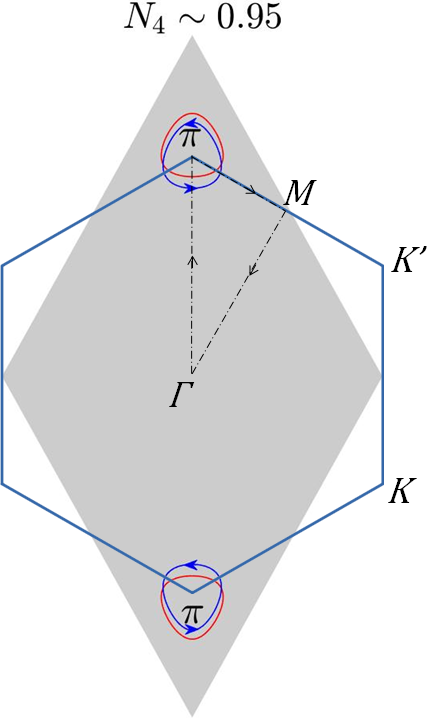}
       \caption{\small{Topological metal arising at non-integer filling fractions. The non-interacting Fermi surface (red line) is compared to the interacting one for $V_1\sim 0.5t$ and $V_2\sim 0.2t$ and three different fillings between $f=1/2$ and $f=2/3$. 
       %The Fermi contour $FC$ is deformed by the interaction while the area of the enclosed surface ($N_4$) stays invariant as is expected. 
       The Berry phase of the partially filled fourth band, $\gamma_4$, is shown together with the filling of the fourth band, $N_4$. The Fermi surface evolves from a single particle-like loop enclosing the QBCP at the $\Gamma$-point and with Berry phase: $\gamma_4\sim  -2\pi$ to two hole-like loops around the Dirac points at $K$ and $K'$, with Berry phases: $\gamma_4^{(K)}=\gamma_4^{(K')}\sim \pi$ giving a total Berry phase of $\gamma_4\sim 2\pi$ associated with this band. The change in the Fermi surface occurs for $N_4 \gtrsim 0.67$}}
 \label{surfaces}
 \end{figure}

In Fig. \ref{surfaces} we show the Fermi surfaces (FS) corresponding to 
three different fillings between $f=1/2$ and $f=2/3$ for $V_1\sim 0.5$ and $V_2\sim 0.2$. For these parameters, the system is deep in the QAH phase as shown in Fig. \ref{fig:fig1}. We compare the interacting Fermi surface with the non-interacting one corroborating Luttinger's theorem which states that the area enclosed by the FS should not vary as we increase the Coulomb interaction. Luttinger's theorem should be satisfied since our calculations are based on a mean-field treatment of the Coulomb interaction. 
 However, the FS can be deformed by the interaction due to the non-local nature of the Fock contribution. A similar deformation of the FS is obtained from the HF approximation to the self-energy, $\Sigma(\omega, {\bf k})$, but on the square lattice.\cite{Valenzuela}
 
 The topological properties of the metallic state generated from the QAH can be investigated by computing the Berry phase associated with the partially filled fourth band \eqref{berry}, $\gamma_4$. As shown in Fig. \ref{surfaces}, by weakly doping the half-filled system with 
 electrons, the FS consists on a single particle-like loop around the $\Gamma$-point.
 The area of the FS coincides with the number of doped electrons in the system off half-filling quantified by the filling of the fourth band: $0 \leq N_4 \leq 1$ . Since the spontaneously broken TRS ground state is two-fold degenerate, we have two possible directions of the chiral currents. For the 
 chiral currents shown in Fig. \ref{scheme}\color{blue}(b)\color{black}, the Berry phase is approximately, $\gamma_4 \sim-2\pi$. As the Fermi level is further increased injecting more electrons in the system, the area of the FS increases as expected, with the same Berry phase as shown in Fig. \ref{surfaces}\color{blue}(a)\color{black}. However, above $N_4\sim 0.67$ the FS is splitted into two hole-like loops around the Dirac points at $K$ and $K'$ Fig. \ref{surfaces}\color{blue}(c)\color{black}. Precisely at this point the total Berry phase of the partially filled fourth band, $\gamma_4$, changes sign and becomes $\gamma_4\sim 2\pi$. This is because each loop around $K$ and $K'$ contributes the same positive Berry phase: $\gamma_4^{(K)}=\gamma_4^{(K')}\sim+\pi$. This also explains why the Chern number of the $n=4$ band is $\nu_4=0$ as shown in Fig. \ref{fig:fig0}.
 
We have followed the same procedure for analyzing the emergence of topological metallic phases in the range $f=2/3\rightarrow 5/6$. Up to an occupancy of the fifth band of $N_5\sim 0.18$, the FS consists of two electron-like loops around $K$ and $K'$. The FS becomes a single hole-like closed loop around the $\Gamma$-point at larger doping. 
Increasing the electron doping further, the area of this loop decreases until it disappears when the fifth band becomes completely filled; at this point the hole occupancy of the fifth band:  $N_5^{(h)}=1-N_5 \rightarrow 0$). At low dopings the Berry phase of each loop 
has the same value but opposite to the case discussed previously: $\gamma_5^{(K)}=\gamma_5^{(K')}\sim -\pi$ until $N_5\sim 0.18$ where the Berry phase 
changes from $\gamma_5\sim -2\pi$ to $\gamma_5\sim 2\pi$. Note that the 
Berry phases obtained are close to but not exactly $2\pi$, and contribute to the non-quantized part of the intrinsic Hall conductivity\cite{Haldane}, $\sigma_{xy}$, as discussed below.

\section{Discussion}
\label{sec:discussion}

We now discuss the implications of our results on experiments. The intrinsic Hall conductivity reads:\cite{xiao}
\begin{equation}
\sigma_{xy}={e^2 \over  \hbar} {1 \over N V} \sum_{n,{\bf k}} f(\epsilon_n({\bf k})) F_n({\bf k})
\end{equation}
where $F_n({\bf k})$ is the Berry curvature of band $n$ and $N$ the number of unit cells with 
volume $V$. In 2D metallic systems $\sigma_{xy}$ can be splitted in two contributions:\cite{Haldane} 
\begin{equation}
\sigma_{xy}=-\frac{e^2}{h} \sum_{n=1}^{N_c}\nu_n- \frac{e^2}{h} {\gamma_{N_c+1} \over 2 \pi}
\label{condformula}
\end{equation}
where $N_c$ is the number of totally occupied bands and $N_c+1$ is the partially filled band. The first term in 
the right hand side of the equation above corresponds to the quantized contribution of 
the $N_c$ occupied bands with corresponding Chern numbers, $\nu_n$. The second term corresponds to the 
non-quantized contribution of the partially filled band which can be expressed 
in terms of the Berry phase, $\gamma_{N_C+1}$, around the Fermi loop enclosing 
the occupied regions in the first Brillouin zone.

\begin{figure}[t!]
   \centering
\includegraphics[width=8.5cm]{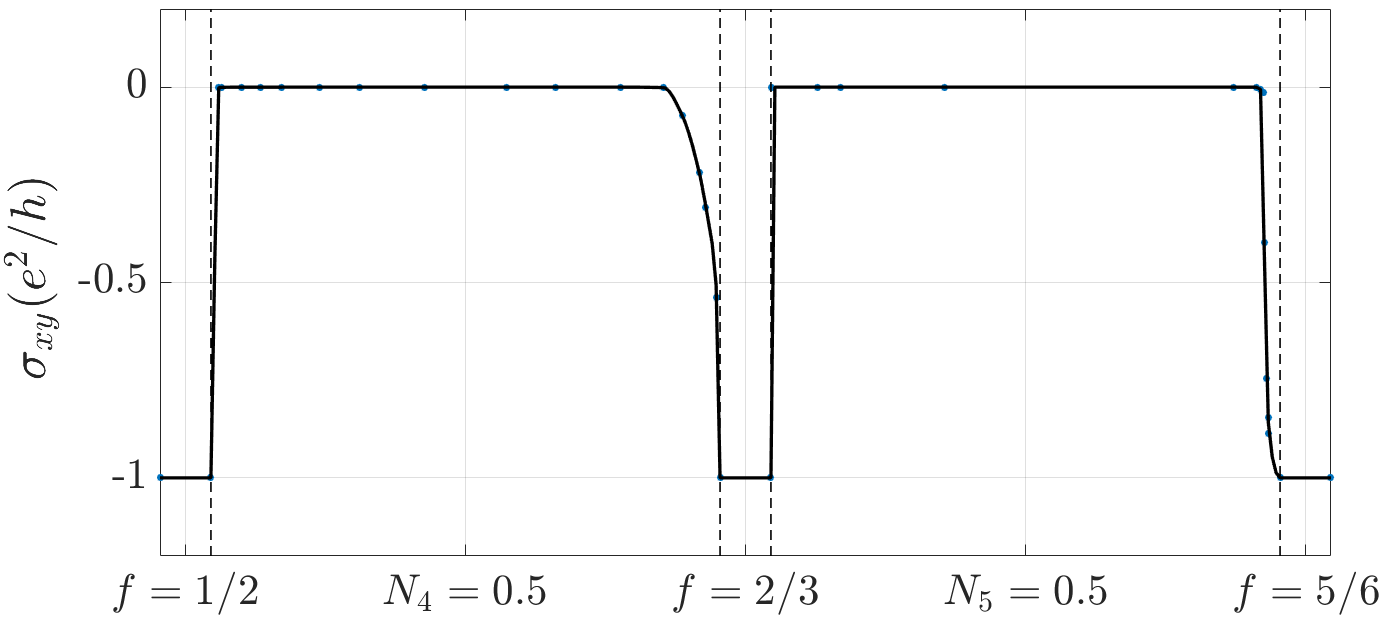}
       \caption{\small{Intrinsic Hall conductivity, $\sigma_{xy}$, as a function of the filling factor, $f$, for $V_1\sim 0.5t$ and $V_2\sim 0.2 t$. In the gapped regions 
      where $f=1/2$ and $f=2/3$, $\sigma_{xy}$ exhibits a plateau, $\sigma_{xy}=-e^2/h$. The sharp but continuous changes in $\sigma_{xy}$ around $f=2/3$ 
     and $f=5/6$ come from the contribution of the fourth or fifth 
     partially filled bands, respectively.}}
 \label{conductivity}
 \end{figure}

In Fig. \ref{conductivity} we show the dependence of the Hall conductivity with electron doping. For instance, at half-filling, we have that $N_c=3$, and electron doping 
is quantified through the filling of the fourth band given by $N_4$. Since the two lowest bands have opposite Chern numbers and the third band has $\nu_3=+1$ 
(see Fig. \ref{fig:fig0}), the half-filled system, $f=1/2$, has a net Hall conductivity of $\sigma_{xy}=-\frac{e^2}{h}$ as shown in Fig. \ref{conductivity}. 
As the fourth band becomes gradually filled for $N_4 < 0.67$, the system becomes metallic with a closed Fermi surface around the 
$\Gamma$-point as shown in Fig. \ref{surfaces}. The Berry phase around such Fermi loop, $\gamma_4\sim -2 \pi$, leads 
to a near complete cancellation with the contribution from the filled bands in Eq. (\ref{condformula}) so the total Hall conductivity, $\sigma_{xy}\sim 0$.

The transition from $\sigma_{xy}=-\frac{e^2}{h}$ to $\sigma_{xy}\sim 0$ is rather abrupt but continuous. The sharpness of the transition is 
related to the small gap existing between the third and fourth bands. If we keep increasing the Fermi energy a sudden change of the 
Fermi surface occurs for $N_4=0.67$ which now consists on two hole-like closed loops around K and K', as shown in Fig. \ref{surfaces}. The 
contribution from the fourth band is still $e^2/h$ leading to $\sigma_{xy}\sim 0$. Hence, in spite of the change
in the Fermi surface, $\sigma_{xy}$ displays a smooth behavior around $N_4 \sim 0.67$.  The Berry phases of the
Fermi loops surrounding K and K' shown in Fig. 5 are {\it both} equal to +$\pi$ which is in contrast to the $\pm \pi$ Berry phases found in 
graphene. Hence, a topological metal characterized by equal Berry phases around the Dirac cones emerges. As the fourth band becomes 
completely filled, $N_4 \rightarrow 1$, we recover $\sigma_{xy} \rightarrow -e^2/h$ precisely for $f=2/3$.
The Hall conductivity between $f=2/3$ and $f=5/6$ behaves in a similar fashion as the fifth band
becomes gradually filled {\it i. e.} as $N_5$ increases.

Our results show that topological metals characterized by non-zero Berry phases on the Fermi surface can be
induced by the off-site Coulomb repulsion in the decorated honeycomb lattice. Such metallic states are protected by $C_6$ rotational symmetry since 
the chiral currents induced at the Hartree-Fock level do not break this symmetry.

%This QAH state with an even Chern number ($\nu = 2$) is also found in the transition metal doped graphene induced by 
%Rashba SOC taking place in the weak magnetization regime.\cite{Zhang2012}

It is well-known that the Berry phase can manifest itself in metals through magnetic oscillatory phenomena.\cite{xiao,mikitik,kopelevich,wright} Semi-classical quantization of electron energy levels leads to the magnetoresistance:
\begin{equation}
\Delta R_{xx} \propto \cos\left[2\pi ({B_F \over B} + {1 \over 2} + \gamma )\right],
\label{eq:magnet}
\end{equation}
where $B_F$ is the frequency of the oscillation
associated with the area of the electron orbit and $\gamma$ the Berry phase (in units of $2 \pi$) picked up by an electron when going around it. For instance, using these measurements,
a Berry phase of $\gamma=1/2$ around the Dirac cone has been obtained in graphene.\cite{kim} Electrons in our topological metal with the FS with closed loops 
around K and K' as shown in Fig. \ref{surfaces}
would also lead to a Berry phase shift of $\gamma=1/2$ in magnetoresistance oscillation experiments. In contrast, the topological metals consisting of a single closed loop FS around the $\Gamma$-point shown in Fig. \ref{surfaces} would have $\gamma \rightarrow 1$. 
This case would be essentially indistinguishable from a topologically trivial metal since it
corresponds to an overall shift of $2 \pi$ in the magnetoresistance oscillations described
by (\ref{eq:magnet}).

\section{Conclusions}
\label{sec:concl}
In the present work we have analyzed, at the mean-field level, the stability of the topologically
non-trivial QAH phase induced by offsite Coulomb repulsion on a decorated honeycomb lattice at different fillings and temperatures. The QAH phase 
occurs when band touching points between the non-interacting valence and 
conduction bands exist. Since the band structure of the decorated honeycomb lattice contains both QBCP and Dirac band touching points it allows analyzing different topological states arising from them. This is interesting since
Dirac points have Berry phases of $\pm \pi$ associated with them while the QBCP have a Berry phase of $\pm 2\pi$\cite{Sun}. In the presence of the Coulomb repulsion a gap can open up around these band touching points. This 
occurs not only at $f=1/2$, as previously reported, but also at $f=2/3$ and $f=5/6$ as shown in Fig. \ref{fig:fig1}.

An important conclusion derived from our study is that the QAH is found to be most stable at $f=1/2$. Also the QAH is more favorable at $f=5/6$ than at $f=1/6$ in spite of the weak Coulomb repulsion acting between the particles in the two cases. 
The QAH is not directly correlated to the existence of QBCP's 
in the non-interacting bands, since we have also found a QAH phase at $f=2/3$ which involves Dirac points. However, the parameter range of $(V_1,V_2)$ in which the QAH phase is stable at $f=2/3$ is smaller than for $f=1/2$. We believe this is due to 
the larger effect of charge frustration due to $ V_2$ at $f=1/2$ as well as the singular behavior of the non-interacting DOS at the Fermi energy associated with the flat band involved in the QBCP. 

The QAH is a topological state protected by $C_6$ rotational symmetry and spontaneously broken TRS which arises in our mean field treatment due to non-zero imaginary Fock amplitudes. The QAH phase competes with the CDW's and NI which are 
topologically trivial insulating phases with the lower reflection and $C_3$ symmetries, respectively instead of the higher six-fold symmetry. It is even possible to find other charge ordering patterns.\cite{24sites} 
%had we considered larger unit cells. For instance, adjoining four unit cell, a $Stripe$ phase with $C_2$ rotational symmetry emerges for $V_1\sim V_2 \gg t$ \cite{24sites}.
%At $f=2/3$ the Dirac points at $K$ and $K'$ with a Berry phase $\gamma^{(K)}=\gamma^{(K')}=+\pi$ are the origin of this topological phase along with the strong enough Coulomb interaction. Nevertheless, at $f=5/6$, the QBCP at $\Gamma$ with a Berry phase $\gamma^{(\Gamma)}=2\pi$ and the singularity in $DOS$ (FIG.\ref{fig:fig0}).   \\
Based on our mean-field treatment we have also analyzed the stability of the QAH phase with temperature finding that the QAH phase at $f=1/2$ is robust up to $T\sim 84$K for the optimum choice of Coulomb parameters: $V_1\sim V_2\sim 1.2t$. This 
temperature can be taken as an overestimation of the actual critical temperature since it is based on a mean-field decoupling of the Coulomb interaction. 

We have finally addressed the question of whether a topological metal can be induced
by doping the QAH phase. We have explored this by injecting electrons in the $f=1/2$ system for Coulomb parameters well in the QAH phase. Interestingly the gaps associated with the QAH phase remain when the system is electron doped between $f=1/2$ and $f=2/3$ and between $f=2/3$ and $f=5/6$. This means that the chiral currents giving rise to the QAH are robust also at such partial filling fractions.
Due to these chiral currents, a topological metal with a non-zero intrinsic Hall conductivity emerges particularly when: $f \lesssim 2/3$ and $f \lesssim 5/6$. Another signature of a topological 
metal arises from the non-zero Berry phases of $\pi$ around K and K' which could, in principle, 
be detected through magnetic oscillatory experiments. 

Future work should include the spin degeneracy in the model and go beyond the present mean-field treatment taking into account electronic correlations in order to check whether the QAH phase and the topological metal found here remain stable. Recent work on the spinfull half-filled Hubbard model \cite{cdmft} (no offsite Coulomb repulsion) on the decorated honeycomb lattice finds a Mott insulator transition between a semimetallic phase and an antiferromagnetic insulator going through an unconventional nematic metallic phase. Both the semimetallic and nematic phases are examples of non-trivial phases emerging 
from the Coulomb repulsion deserving further characterization. 
%Since we carry out a Hartree-Fock mean field treatment, this gives us an estimation of the temperature at which QAH can be detected on DHC lattices.
%We have shown how the QAH state with $C_6$ invariance, when one band is partially filled, this is in a metallic regime.
\acknowledgments
We acknowledge financial support from (MAT2015-66128-R) MINECO/FEDER, Uni\'on Europea.

\appendix 
\section{Berry phases and Chern numbers in multiband systems}
 
In a discretized Brillouin zone, the total Chern number $\nu$ can be calculated from the Berry phases $\gamma_{nl}$ at each elementary placquette $l$. The Berry phase is just the accumulated phase of the wave function along a certain closed $k$-path. 
\begin{equation}
\gamma_{nl}=\Im \ln\prod_{j=0}^{N-1}\bra{u_{\bf{nk}_j}}\ket{u_{\bf{nk}_{j+1}}}
\end{equation}
where $n$ is the band index and the loop chosen is rectangular with $N=4$. In the multiband case the wave functions overlap of all possible combinations must be taken into account. So that, if we have $N_{c}\equiv 6f$ valence bands, we construct a $N_{c}\times N_{c}$ matrix at each step of the path. Then, the Berry phase is just the phase of the determinant of the product of these matrices along the loop. The Chern number is nothing else than the sum over the $FBZ$ of all those Berry phases:
\begin{equation}
\nu=\frac{1}{2\pi }\sum_{FBZ} \Im\ln\det\prod_{j=0}^{3}\bra{u_{\bf{mk}_j}}\ket{u_{\bf{nk}_{j+1}}}
\label{ChernForm}
\end{equation}
where $1\leq m,n \leq N_c$. The shortest steps the closer to an entire Chern number ($\nu\in \mathbb{Z}$).\cite{Fukuki} If the band is partially filled, for instance by doping the material, the system turns into a conductor. If the Fermi surface (FS) is a simple closed loop, the Berry phase of the metal is determined at this path. Then in \eqref{ChernForm} we restrict the sum to the enclosed surface (FS):
\begin{equation}
\gamma_n=\frac{1}{2\pi }\sum_{FS} \Im\ln\prod_{j=0}^{3}\bra{u_{\bf{nk}_j}}\ket{u_{\bf{nk}_{j+1}}}
\label{berry}
\end{equation}
where $n$ corresponds to the partially filled band. This Bery phase do not have to be a multiple of $2\pi$ contributing to the nonquantized part of the intrinsic Hall conductivity $\sigma_{xy}$.\cite{Haldane}\\

\section{Stability analysis of the ground state}
In the mean-field treatment used in the paper we find that for a given set of parameters ($V_1,V_2$), there are several solutions having very close free energies. 
\begin{figure}[h]
\includegraphics[width=4.18cm]{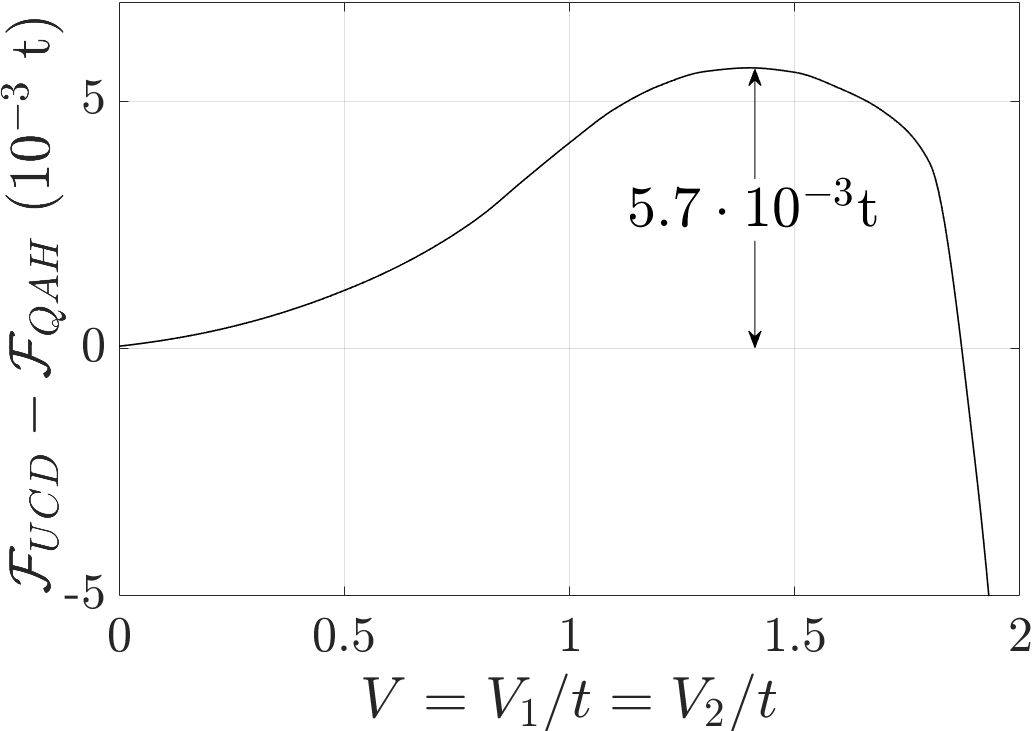}
\includegraphics[width=4.32cm]{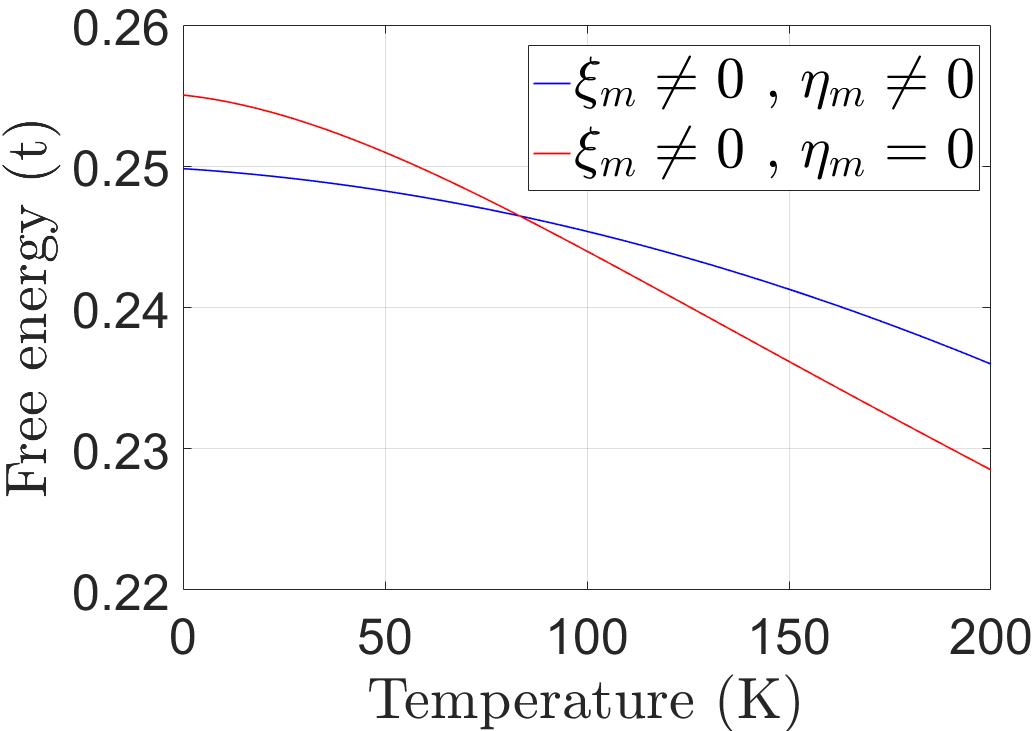}
       \caption{\small{(a) Dependence of the free energy difference between the two lowest energy states for $V_1/t=V_2/t\equiv V$ and $f=1/2$. The lowest free energies are found for
       the UCD with $\eta_m\equiv 0$ and the QAH with $\eta_m\neq 0$. In the range $0\lesssim V\lesssim 1.9$, the free energy difference is positive meaning that the QAH phase is the ground state. It presents a maximum at $V\sim 1.4$. (b) The UCD and QAH energies as a function of $T$. The QAH is the ground state up to $T\sim 84 K$ where the UCD becomes the lowest energy phase. }}
 \label{comparison}
 \end{figure}
 
In Fig. \ref{comparison}{\color{blue}(a)} we show the free energy difference between the UCD and the QAH at $f=1/2$ for different $V_1/t=V_2/t\equiv V$ in the $T \rightarrow 0$ limit. At $V\sim1.4$ the QAH is the most stable with an energy of $0.0057 t$ below the UCD. This is the maximum energy difference between the UCD and the QAH as $V$ is increased which is consistent with the maximum $T$ at which the QAH phase 
survives around $V\sim 1.2$ as shown in Fig. \ref{VT}.
In Fig. \ref{comparison}{\color{blue}(b)} we show the dependence on temperature on the two lowest energy states at fixed $V_1\sim V_2\sim 1.2t$. This plot shows how for $T \lesssim 84$ K the QAH has the lowest energy but for $T>84$ K the UCD becomes the ground state. Hence, $T \sim 84$ K is the maximum temperature estimate at which the QAH is stable as shown in Fig. \ref{VT}.

\end{document}